\documentstyle[preprint,revtex]{aps}
\begin{document}
\draft
\begin{title}
One Dimensional Oxygen Ordering in YBa$_2$Cu$_3$O$_{7-\delta}$
\end{title}
\author{A.A. Aligia}
\begin{instit}
Comisi\'{o}n Nacional de Energ\'{\i}a At\'{o}mica \\
Centro At\'{o}mico Bariloche e Instituto Balseiro\\ 8400 S.C. de Bariloche,
R\'{\i}o~Negro,
Argentina
\end{instit}
\begin{abstract}
A model consisting of oxygen-occupied and -vacant chains is
considered, with repulsive first and second nearest-neighbor
interactions $V_1$ and $V_2$, respectively.  The statistical
mechanics and the diffraction spectrum of the model is solved
exactly and analytically with the only assumption $V_1 >> V_2$.  At
temperatures $T \sim V_1$ only a broad maximum at (1/2,0,0) is
present, while for $\mid \delta - 1/2 \mid > 1/14$ at low enough
$T$, the peak splits into two.  The simple expression for the
diffraction intensity obtained for $T << V_1$ represents in a more
compact form previous results of Khachaturyan and Morris, extends
them to all $\delta$ and $T/V_2$ and leads to a good agreement with
experiment.
\end{abstract}
\pagebreak
\par It is now clear that the superconducting transition
temperature of YBa$_2$Cu$_3$O$_{7-\delta}$ depends not only on
$\delta$ but also on the oxygen ordering \cite{veal90}.  Therefore,
detailed knowledge of this ordering in the whole oxygen
concentration range is important for an understanding of the
electronic properties of the system and the superconducting mechanism.
\par At low enough temperatures,
at least for $1/4 < \delta \leq 1/2$, the oxygen atoms in the
basal planes are ordered in infinite CuO chains \cite{bey89,rey89}.
 Experimental \cite{rey89,son91,son92,al88} and theoretical
\cite{aligia92,q92,seme,agb} evidence favoring alternative types of
oxygen ordering is restricted to $\delta > 0.6$
and $\delta \leq 0.25$.  Concerning the ordering among different
CuO chains at $T=0$,
the experimental situation \cite{bey89,rey89} favors
structures that are given by the ground state of a one-dimensional
(1D) Ising model in which the interactions $V_n$ between  chains at
a distance $na$, where $a$ is the lattice parameter, satisfy the
inequality $V_{n+1} + V_{n-1} > 2V_n$
\cite{agb,g1d,aligia91,ceder90}.  The low temperature thermodynamics
of the two-dimensional asymmetric next-nearest-neighbor Ising model
(ASYNNNI) \cite{wille88} is also governed by the simplest
one-dimensional Ising model ($V_1 >0$ and $V_n$ = 0 for $n > 1)$
\cite{ceder90,font89}.
\par Due to the sluggish oxygen kinetics at low temperatures, the
system often does not reach a completely ordered state at $T$=0, as
is required by the third principle of thermodynamics.  In this
case, in the range $1/4 \leq \delta \leq 1/2$, diffuse
diffraction peaks are observed \cite{bey89,lev92}.  For the largest
values of $\delta$, and diffraction vectors {\bf q} = ($2 \pi/a)(h, 0,0$),
only one peak in the interval 0$<h<$1,centered at $h$ = 1/2 is present.
For the smallest values of $\delta$ in the above mentioned range, two
maxima centered at $h = 1/2 \pm \epsilon$ are observed.
Khachaturyan and Morris (KM) \cite{khacha90} explained
qualitatively these observations in the range $1/3 \leq \delta \leq
1/2$, assuming a random faulting of the double period ordered
structure with $\delta=1/2$ (alternating Cu-O and Cu-vacancy chains
\cite{rey89}).
This work has been criticized in a Comment \cite{comm} because of the
restricted composition range of the theory
and the assumptions on the interactions
that would be implicit in the model. In their Reply \cite {rep} KM state
that Ref. \cite{khacha90} does not assume a particular type of
interaction and that, while it would be nice to have a tractable
analysis for all $\delta$, it was not necessary to establish that
random faulting can produce split peaks. In another Comment to
\cite{khacha90}, it was
shown that the short-range order
implicitly assumed by KM, minimizes the free energy of the 1D Ising
model with repulsive $V_1$, $V_2$ and $V_n$=0 for $n>2$ at $T
\rightarrow 0$ \cite{aligia90}.  $V_1 > 2V_2$ should hold to
stabilize the double period structure at $\delta = 1/2$.
While interactions at larger distances are important in determining
the ground state of the system \cite{agb,g1d,aligia91,ceder90}, it
will be shown in this letter that this model is the simplest one which
leads to a reasonable agreement with the experimentally observed
diffraction peaks.
However,
for $T \rightarrow 0$ the free energy should be minimized by
long-range ordered structures and this model becomes unrealistic.
In addition, the results of KM predict too narrow and intense split
peaks near $\delta = 1/3$ when compared with experiment
\cite{bey89,lev92}.  Since one expects that these peaks should
broaden and lose intensity when the temperature is
increased, it is of interest not only to extend the theory of KM to all
compositions, but also to study the model at finite temperatures.
This task is carried out in the present letter.
\par In
order to obtain simple analytical results, two cases are
considered: a) $T >> V_2$, $V_1/T$ arbitrary \cite{foot},
b) $V_1 >> T$ and any
$V_2/T$ with $V_2 < V_1/2$.  The resulting short-range correlations
are used as input parameters for the calculation of the diffraction
intensity.  If the system is metastable, these parameters can be
thought of as probabilities that are given by the preparation method,
independent of the free energy of the system, or as equilibrium
parameters at higher $T$.  While KM obtained two different
expressions for the diffraction intensity, one for $1/3 \leq \delta
\leq 4/9$ and another one for $4/9 \leq \delta \leq 1/2$
\cite{khacha90}, in case (b) the elegant method based on
generating functions \cite{schwa83} leads to a single expression
which simplifies those of KM, extends them to all oxygen
compositions, and allows for more than two consecutive oxygen
occupied chains (requiring $T \neq0$ in the 1D Ising model).
\par Following KM we shall denote by $\bigcirc$ the Cu-O chains and
by $\Box$ the Cu-vacancy chains.  It is convenient to write the
model in the form:
\begin{equation}
H = V_1 \sum_i n_i n_{i+1} + V_2 \sum_i n_i n_{i+2},
\end{equation}
where $n_i =1$ (0) if the ith chain is $\Box (\bigcirc$).  The
diffraction intensity is given by \cite{khacha90}:
\begin{equation}
I (h) = N \mid f_{ox} \mid^2 \delta \sum^{\infty}_{m=- \infty} P(m)
\exp (i2 \pi hm),
\end{equation}
where $N$ is the total number of unit cells, $f_{ox}$ is the oxygen
scattering factor and $P(m)$ is the conditional probability that if
$n_0=1$, also $n_m=1$.  Interchanging $\bigcirc$ and $\Box$ it is
easily seen that, excluding integer $h$, $I(h)$ is the same for
$\delta$ and $1- \delta$.
\par In the following $\delta \leq 1/2$ is assumed.
\par
\noindent a) $V_2 << T$
\par In this case $V_2$ can be neglected (as will become clearer in
case (b)) and the model reduces to the simplest 1D Ising model.  This
model describes the low-temperature oxygen ordering of the 2D
ASYNNNI model \cite{ceder90,font89} and is exactly solvable
\cite{kiku51}.  The quantity that determines the free energy and
$I(h)$ for each value of $\delta$, is the probability $y_1$ of
finding a pair $\Box \Box$ of two consecutive $\Box$ chains:
\begin{equation}
y_1 = \delta - \gamma_1/2 + \left [ (\delta - \gamma_1/2)^2 +
\delta^2 (\gamma_1 -1) \right ]^{1/2},
\end{equation}
where $\gamma_1 = [1 -\exp (-V_1/T)]^{-1}$.
$P(m)$ satisfies the following non-homogeneous difference equation:
\begin{equation}
P(m) = \eta - \beta P(m-1),
\end{equation}
where $\eta = (\delta- y_1)/(1-\delta)$ is the conditional
probability that if $n_{m-1}=0$, then $n_m =1$.  Similarly $\eta -
\beta = y_1/\delta = P(1)$ is the probability that if $n_{m-1}=1$
then $n_m=1$.  The solution of Eq. (4) with the boundary condition
$P(0)=1$ is:
\begin{equation}
P(m) = \delta + (1-\delta)(-\beta)^{\mid m \mid},
\end{equation}
and replacing this in Eq.~(2)
\begin{equation}
I(h) = N \mid f_{ox} \mid^2 \delta \frac {(1-\beta)(1+\beta -
\eta)}{1+ \beta^2 + 2 \beta \cos (2\pi h)}.
\end{equation}
For any $\delta$ and $T$ this expression gives only one peak
centered at $h=1/2$.  If enough statistics and Monte Carlo steps
are allowed, Monte Carlo results using the ASYNNNI \cite{font89} at
low enough temperatures should converge to this simple expression.
Thus, the ASYNNNI should be extended to include longer range
repulsions in order to explain split diffraction peaks
\cite{agb,g1d,aligia91,ceder90}, as shown before \cite{aligia90}.
\par In the limit $T >> V_1$ \cite {foot},
then $\gamma_1 = T/V_1,~ y_1 =
\delta^2 [1 - (1- \delta)^2 V_1/T],~ \eta = \delta + \delta^2 (1-
\delta) V_1/T,~ \beta = \delta(1- \delta) V_1/T$ and:
\begin{equation}
I(h) \stackrel {\sim}{=} N \mid f_{ox} \mid^2 \delta (1-\delta)
\left [ 1 - 2 \delta (1- \delta) (V_1/T) \cos (2 \pi h) \right ].
\end{equation}
Thus, at high enough $T,~I(h)$ is a constant plus a small harmonic
term with maximum at $h = 1/2$ and minimum at $h \rightarrow 0$ and
$h \rightarrow 1$.
\par For $V_1 >> T >> V_2$, neglecting exponentially small terms,
$\gamma_1 = 1$, $y_1 = 0$, $\eta = \beta = \delta/ (1- \delta)$ and:
\begin{equation}
I(h) = \frac {N \mid f_{ox} \mid^2 \delta (1- \delta) (1-2
\delta)}{1 +2 \delta (1- \delta) \cos (2 \pi h)},
\end{equation}
in agreement with Ref. \cite{aligia90}. For $\delta \rightarrow 0,~I(h)$ is
small and flat, while for $\delta \rightarrow 1/2$, the second
member of Eq.~(9) gives the extremely narrow peak at $h=1/2$
corresponding to the double period long-range ordered structure.
\par
\noindent b) $V_1 >> T$
\par As shown above, for $\delta \leq 1/2$ the probability $y_1$ of
finding a strip $\Box \Box$ is of order $\delta^2 (1-2 \delta)^{-1}$
exp ($-V_1/T$) and can be neglected if $\delta \neq 1/2$.  This
allows to represent any possible structure in terms of a sequence
of two strips: $\bigcirc$ and $\Box \bigcirc$.  The energy per
strip is given by $V_2 y$ where $y$ is the probability of finding two
nearest-neighbors $\Box \bigcirc$ strips.  Calling $x$ the
probability of finding one $\Box \bigcirc$ strip, it is easy to see
that the problem can be mapped into the 1D Ising model with only
$V_1 \neq 0$ ,already considered in case (a).  The mapping is given
by the correspondence $\Box \bigcirc \rightarrow \Box$,$x
\rightarrow \delta$, $y \rightarrow y_1$, $V_2 \rightarrow V_1$.
The free energy per chain $F$ is given by:
\begin{eqnarray}
(1+x)F = V_2 y -T [ x \ln x + (1-x)\ln (1-x) - y\ln y
- 2 (x-y) \ln (x-y) \nonumber \\
- (1-2x+y) \ln (1-2x+y) ],
\end{eqnarray}
where $1+x$ is the average number of chains per strip.  The average
number of $\Box$ chains per strip is $x/(1+x) = \delta$.  Thus:
\begin{equation}
x = \delta/ (1- \delta),
\end{equation}
and minimizing $F$ one obtains:
\begin{equation}
y = x - \gamma/2 + \left [(x- \gamma/2)^2 + x^2 (\gamma -1) \right ]^{1/2},
\end{equation}
where
\begin{equation}
\gamma = \left [ 1 - \exp (-V_2/T) \right ]^{-1}.
\end{equation}
The probability of finding three consecutive $\bigcirc$ chains, which is
given by $1+ \delta (y/x - 3)$ can be used instead of $T$ as
independent variable.
\par $P(m)$ can be determined from $P(m-1)$ and $P(m-2)$ from the
following reasoning. Since $P(m)=P(-m), m \geq 0$ will be assumed.
The pair of chains $m-2$ and $m-1$ can be in
one of the three following states: 1) $\bigcirc \bigcirc$, 2) $\Box
\bigcirc$, 3) $\bigcirc \Box$.  The probability of state $i$ is
denoted $p_i$.  Since $\Box \Box$ is excluded, if the third state
is realized ($p_3$=1) then $P(m)$=0.  In state 1,
the chain $m-1$ should belong to a $\bigcirc$
strip and if $p_1$ =1, $P(m)$ is given by the
conditional probability $(x-y)/(1-x)$ that after a strip $\bigcirc$,
a strip $\Box \bigcirc$ follows.  Similarly, if $p_2$ = 1, $P(m)$
is the conditional probability $y/x$ that after a strip $\Box
\bigcirc$, another of the same kind follows.  Thus:
\begin{equation}
P(m) = p_1 (x-y)/(1-x) + p_2 y/x,
\end{equation}
and using $P(m-2)=p_2$,~ $P(m-1)=p_3$,~ $p_1 +p_2 +p_3 = 1$, Eq.~(13)
takes the form:
\begin{equation}
P(m) = \beta \left [1 - P(m-1) \right ] - \alpha P(m-2),
\end{equation}
where
\begin{equation}
\beta = (x-y)/(1-x)~;~ \alpha = \beta -y/x.
\end{equation}
Eq.~(14) can be solved using the generating function \cite{schwa83}:
\begin{equation}
G(z) = \sum^{\infty}_{m=0} P(m) z^m.
\end{equation}
Using Eq.~(14) and the boundary conditions $P(0)=1$, $P(1)=0$, an
equation for $G(z)$ is obtained, the solution of which reads:
\begin{equation}
G(z) = \frac{1+(\beta-1)z}{(1-z)(1+ \beta z + \alpha z^2)}.
\end{equation}
By integration in the complex plane it can be shown that
\begin{equation}
P(m) = \delta - R_{m+1} (z_1) - R_{m+1} (z_2),
\end{equation}
where $R_n(z_i)$ is the residue of $G(z)/z^n$ at the  pole $z_i$,
and $z_1$ and $z_2$ are the two roots of the polynomial $\alpha z^2 +
\beta z + 1$.  However, the diffraction intensity can be obtained
directly from the generating function.  Using Eqs. (2) and
(16) one has:
\begin{equation}
I(h) = N \mid f_{ox} \mid^2 \delta \left [ G (\exp (i 2 \pi h)) +
G(\exp (-i2 \pi h)) -1 \right ],
\end{equation}
and after some algebra, this expression is simplified to:
\begin{equation}
I(h) = \frac {N \mid f_{ox} \mid^2 \delta (1- \alpha)(1+ \alpha -
\beta)} {4 \alpha \cos^2 (2 \pi h) + 2 \beta(1+ \alpha) \cos (2 \pi h)
+ (1- \alpha)^2 + \beta^2}
\end{equation}
Eq.~(20), together with Eqs. (10) to (12) and (15) describes the
scattering intensity for all $\delta \leq 1/2$ and $T<< V_1$.  For
$\delta > 1/2,~ I(h)$ is given by the same equation with $\delta$
replaced by $1- \delta$.  The condition for the existence of two
split maxima is obtained requiring that the denominator of Eq.~(20) as
a function of $\cos(2 \pi h)$ has a minimum in the interval (-1,1).
One obtains $4 \alpha > \beta (1+ \alpha)$, or equivalently $\gamma
< \gamma_c$, where
\begin{equation}
\gamma_c = 4 (1-2 \delta)/(1- \delta)
\end{equation}
Since $\gamma \geq 1$, split peaks are possible only for $\delta <
3/7 = 0.429$.  For these values of $\delta$, the simple Eqs.~(12)
and (21) give the critical temperature above which only one peak
exists. In the region of compositions and temperatures for which
two intensity maxima exist ($\delta < 3/7$ and $\gamma < \gamma_c$),
their positions are given by a very simple expression:
\begin{equation}
h_{max} = \frac {1}{2 \pi} \arccos \left [ - \frac {\gamma (1-
\delta)} {4(1-2\delta)} \right].
\end{equation}
\par In the limit $V_1 >> T >> V_2$, neglecting terms of order $\exp
(-V_1/T)$ and of order $V_2/T$, which do not bring any qualitative
change in $I(h)$, one has $\gamma = T/V_2,~ y = x^2,~ \beta = x,~
\alpha = 0$ and $I(h)$ takes the form of
Eq.~(8).  For $T<<V_2$, neglecting exponentially small terms one
has $\gamma=1$ and i) for $\delta \leq 1/3$,~ $y = 0,~ \alpha = \beta =
\delta/(1-2 \delta)$; ii) for $\delta \geq 1/3$,~ $y = (3
\delta-1)/(1- \delta),~ \beta =1,~ \alpha = (1-2 \delta)/ \delta$.
Case (ii) coincides with the one previously solved by KM
\cite{khacha90}, since Eq.~(14) takes the form of the
nonhomogeneous difference equation of KM.  In {\it both} cases,
$I(h)$ takes the form (for $T<<V_2$ and any $\delta$):
\begin{equation}
I(h) = \frac {N \mid f_{ox} \mid^2 \delta (1-2 \delta) \mid 1-3
\delta \mid}{4 \delta (1-2 \delta) \cos^2 (2 \pi h) + 2 \delta (1-
\delta) \cos (2 \pi h) + 10 \delta^2 + 1 - 6 \delta}
\end{equation}
Eq.~(10a) (valid for $4/9 \leq \delta \leq 1/2$) and also Eq.~(10b)
of KM ($1/3 \leq \delta < 4/9$) should reduce to the much simpler
Eq.~(23).  Moreover, Eq.~(23) extends the results of KM to all
values of $\delta$, and Eq.~(22) with $\gamma=1$ gives the position
of the split maxima. This low temperature limit can be described as
an uncorrelated sequence of i) strips $\bigcirc$ and
$\Box \bigcirc \bigcirc$ for $\delta \leq 1/3$ or ii) strips
$\Box \bigcirc$ and $\Box \bigcirc \bigcirc$ for $\delta \geq 1/3$.
These strips would become correlated if $V_3$ (and $V_4$ for
$\delta \geq 1/3$) were included in the model.
\par For a comparison with experiment, the
low temperature limit Eq.~(23) is not good enough and Eqs.~(10) to
(12), (15) and (20) should be used.  In Fig.~1, the
evolution of $I(h)$ with temperature is represented.
For $\delta=0.364$ and $T=0$
($\gamma = 1)$, $I(h)$ shows two sharp peaks as already shown in
Fig.~1 of KM \cite{khacha90}.
As expected, the peaks broaden and lose intensity, keeping the same
total area, when the temperature is increased. However, as long as
two well defined maxima exist, the positions of the peaks do not
depend strongly on temperature.
For $\gamma=1.2$,
the result is very similar to one of those obtained by Beyers {\it et
al.} for $\delta = 0.35$ \cite{bey89,lev92}.  For $\delta = 0.25$,
the experimental peaks are somewhat sharper, suggesting that
repulsions at larger distances than two lattice parameters are also
present \cite{agb,g1d,aligia91,ceder90}.  Other difficulties in the
comparison between theory and experiment are the possibility of
phase separation \cite{bey89,lev92} and that a fraction of oxygen
atoms always remains disordered \cite{krek92}, particularly for
quenched samples.  For comparison with the experimental results in
quenched samples \cite{lev92} with $\delta = 0.27$, $\delta = 0.35$
and $\delta = 0.43$, $\delta$ is replaced by $5 \delta/4$ in the
theoretical curves, following Ref. \cite{krek92}.
The corresponding results
for $\gamma = 1.2$, shown in Fig.~2, are in good agreement with
experiment.  The intensity for $\delta = 0.35 \times 5/4$ is
somewhat higher than the experimental one.
\par The agreement with experiment can be improved by adding more
interactions.  Also a quantitative agreement with experiment was
obtained postulating that $P(2n+1) = 0$ within domains, the size of
which is adjusted for each $\delta$ \cite{lev92}.  However, except
in unrealistic limits, an analytical treatment of the problem is no
longer possible in these cases and a further improvement of the
present results can only be done at the cost of a loss in simplicity
and physical transparency. Furthermore, in view of the above mentioned
experimental uncertainties, the analytical results are already
satisfactory.
\par I acknowledge gratefully insightful conversations with
J. Garc\'{e}s and H.Wio.

\figure{Intensity as a function of diffraction vector for several
values of $V_2/T$ (see Eq.~(12)), $V_1 >> T$ and (a) $\delta =
4/11$, (b) $\delta = 1/4$. In (a), the critical temperature above which
only one peak exists is given by $\gamma=\gamma_c=12/7$ (see Eqs.~(12)
and (21))}
\figure{Intensity as a function of diffraction vector for $\gamma =
1/2$, $V_1 >> T$ and several values of $\delta$.}


\begin{references}
\bibitem{veal90}B.W. Veal {\it et al.}, Phys. Rev. B {\bf 42}, 6305 (1990).
\bibitem{bey89}R. Beyers {\it et al.}, Nature (London) {\bf 340}, 619 (1989).
\bibitem{rey89}J. Reyes-Gasga {\it et al.}, Physica C{\bf 159}, 831 (1989).
\bibitem{son91}R. Sonntag {\it et al.}, Phys. Rev. Lett. {\bf 66}, 1497
(1991).
\bibitem{son92} Sonntag, Th. Zeiske and D. Hohlwein, Physica B
{\bf 180-181}, 374 (1992).
\bibitem{al88}M.A. Alario-Franco {\it et al.}, Physica C {\bf 156}, 455
(1988).
\bibitem{aligia92}A.A. Aligia, Europhys. Lett. {\bf 18}, 181 (1992).
\bibitem{q92}Qiang Wang {\it et al.}, Phys. Rev. B {\bf 45}, 10834 (1992).
\bibitem{seme}S. Semenovskaya and A.G. Khachaturyan, Phys. Rev. B
{\bf 46}, 6511 (1992).
\bibitem{agb}A.A.Aligia, J.Garc\'{e}s and H.Bonadeo, Phys.
Rev. B {\bf 42}, 10226 (1990).
\bibitem{g1d}D.de Fontaine, G.Ceder, and M.Asta, Nature (London) {\bf 343},
544 (1990)
\bibitem{aligia91}A.A. Aligia, H. Bonadeo and J. Garc\'{e}s, Phys.
Rev. B {\bf 43}, 542 (1991).
\bibitem{ceder90}G. Ceder {\it et al.}, Phys. Rev. B {\bf 41}, 8698
(1990); references therein
\bibitem{wille88}L.T. Wille, A. Berera and D. de Fontaine, Phys.
Rev. Lett. {\bf 60}, 1065 (1988).
\bibitem{font89}D. de Fontaine, M.E. Mann and F. Ceder, Phys. Rev.
Lett. {\bf 63}, 1300 (1989).
\bibitem{lev92}L.E. Levine and M. D\"{a}umling, Phys. Rev. B {\bf
45}, 8146 (1992).
\bibitem{khacha90}A.G. Khachaturyan and J.W. Morris, Jr., Phys.
Rev. Lett. {\bf 64}, 76 (1990).
\bibitem{comm}D. de Fontaine and S.C. Moss, Phys. Rev. Lett. {\bf 67},
527 (1991).
\bibitem{rep}A.G. Khachaturyan and J.W. Morris, Jr., Phys. Rev. Lett.
{\bf 67}, 528 (1991).
\bibitem{aligia90}A.A. Aligia, Phys. Rev. Lett. {\bf 65}, 2475 (1990).
\bibitem{foot}When the 1D model is regarded as a limiting case of 2D
models of O ordering \cite{wille88,agb}, the temperature
should be low enough to
ensure a correlation length of several lattice parameters along the
CuO chains. A study of the relevant correlation functions suggests
that the limiting temperature lies roughly 30\% below the tetragonal-
orthorhombic transition temperature for $\delta=1/2$
\cite {ceder90,agb}.
\bibitem{schwa83}for example W.A. Schwalm and M.K. Schwalm, Am.
J. Phys. {\bf 51}, 230 (1983).
\bibitem{kiku51}for example Mc Coy, {\it The Two Dimensional Ising Model},
Harvard, Cambridge, Masacusets (1973); R. Kikuchi, Phys. Rev. {\bf 81},
988 (1951).
\bibitem{krek92}T. Krekels {\it et al.}, Physica C {\bf 196}, 363 (1992);
references therein.
\end{references}
\end{document}